\documentclass[prl,amsmath,amssymb,twocolumn, longbibliography]{revtex4-2}
\usepackage{graphicx}
\usepackage{subfigure}
\usepackage{adjustbox}
\usepackage{bm}
\usepackage{color}
\usepackage{braket}
\usepackage{standalone}
\usepackage{multirow}
\usepackage{tikz}
\usepackage{mathrsfs}
\usepackage{dsfont}
\usepackage{comment}
\usepackage{amsmath}
\usepackage[colorlinks,bookmarks=true,citecolor=blue,linkcolor=red,urlcolor=blue]{hyperref}
\usepackage{cleveref}
\usepackage{orcidlink}

\graphicspath{{./figures/}}

\begin{document}

\title{Fingerprints of Composite Fermion Lambda Levels in Scanning Tunneling Microscopy}
\author{Songyang Pu$^{1,2*}$\orcidlink{https://orcid.org/0000-0003-0144-1548}, Ajit C. Balram$^{3,4*}$\orcidlink{0000-0002-8087-6015}, Yuwen Hu$^{5}$, Yen-Chen Tsui$^{5}$, Minhao He$^{5}$, Nicolas Regnault$^{6,5}$, Michael P. Zaletel$^{7}$, Ali Yazdani$^{5}$, Zlatko Papi\'c$^{1}$ }

\affiliation{$^{1}$School of Physics and Astronomy, University of Leeds, Leeds LS2 9JT, United Kingdom}
\affiliation{$^{2}$Department of Physics and Astronomy, The University of Tennessee, Knoxville, TN 37996, USA}
\affiliation{$^{3}$Institute of Mathematical Sciences, CIT Campus, Chennai, 600113, India}

\affiliation{$^{4}$Homi Bhabha National Institute, Training School Complex, Anushaktinagar, Mumbai 400094, India}
\affiliation{$^{5}$Department of Physics, Princeton University, Princeton, NJ 08544, USA}
\affiliation{$^{6}$Laboratoire de Physique de l’Ecole normale sup\'erieure,ENS, Universit\'e PSL, CNRS, Sorbonne Universit\'e}
\affiliation{$^{7}$Department of Physics, University of California at Berkeley, Berkeley, CA 94720, USA}

\date{\today}

\begin{abstract}
Composite fermion (CF) is a topological quasiparticle that emerges from a non-perturbative attachment of vortices to electrons in strongly correlated two-dimensional materials.  Similar to non-interacting fermions that form Landau levels in a magnetic field, CFs can fill analogous ``Lambda'' levels, giving rise to the fractional quantum Hall (FQH) effect of electrons. Here, we show that Lambda levels can be directly visualized through the characteristic peak structure in the signal obtained via spectroscopy with the scanning tunneling microscopy (STM) on a FQH state. Complementary to transport, which probes low-energy properties of CFs, we show that \emph{high-energy} features in STM spectra can be interpreted in terms of Lambda levels. We numerically demonstrate that STM spectra can be accurately modeled using Jain's CF theory. Our results show that STM provides a powerful tool for revealing the anatomy of FQH states and identifying physics beyond the non-interacting CF paradigm. 
\end{abstract}

\maketitle

{\bf \em Introduction.---}Fractional quantum Hall (FQH) phases~\cite{Tsui82} of matter possess nonlocal order which gives rise to topologically-quantized Hall conductance, dissipationless boundary modes, and emergent quasiparticle excitations that are distinct from fermions or bosons~\cite{Laughlin83, Prange87, DasSarma07, JainHalperin2020}. While some of these properties have been successfully accessed via transport~\cite{Tsui82} and interferometry~\cite{Nakamura20, Willett2023} measurements, recent advances in the scanning tunneling microscopy (STM)~\cite{Xiaomeng2022, Coissard2022, Farahi2023, Hu23} have opened a new window to directly probe FQH states at much higher energies than in the past. The sensitivity of the early spectroscopy experiments on GaAs materials~\cite{Ashoori90, eisenstein92b, Dial2007, Dial2010, Yoo19} was heavily constrained by the two-dimensional electron gas (2DEG) residing deep inside the semiconductor heterostructures. These limitations have recently been lifted in two important ways: by utilizing ultra-clean graphene materials, which host FQH states atomically close to the vacuum~\cite{Du09, Bolotin09, Dean11, Feldman12, Kou2014, Ki14, Maher2014, Amet15}, and by STM tip preparation ~\cite{Farahi2023, Hu23} that allows performing non-invasive imaging of FQH states. Moreover, in samples with a few defects, STM was used to directly probe the spatial structure of the Landau orbits~\cite{LuicanMayer2014} and, in materials such as bismuth, it was used to visualize lattice-symmetry broken ground states~\cite{Feldman16}.

\begin{figure}[b]
    \centering    
    \includegraphics[width=\linewidth]{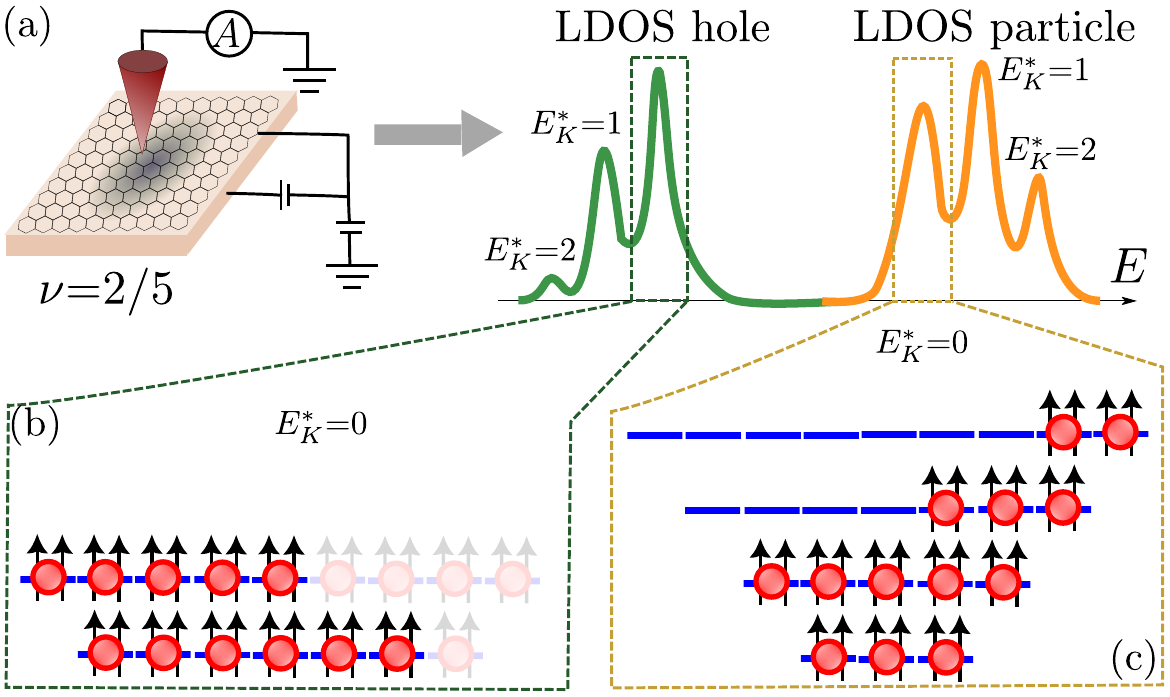}
    \caption{(a) Schematic of the STM probe of a $\nu{=}2/5$ FQH state and measured signal on the hole and particle sides. (b)-(c) The signal peaks are fingerprints of CF Landau levels.
    The first peak on the hole side corresponds to the $E_K^*{=}0$ state where five CF holes are in the lowest two $\Lambda$Ls as shown (see text for the definition of $E_K^*$). On the particle side, the $E_K^*{=}0$ state is obtained when one electron is added, corresponding to five CF particles that occupy higher, $n{\geq}2$, $\Lambda$Ls. 
    }
    \label{fig: sketch}
\end{figure}

In light of these developments, a question arises: what does the STM, performed on a FQH state, actually measure? A textbook answer is that STM probes the Local Density of States (LDOS) for injecting or removing an electron from the system. However, the underlying quasiparticles of  FQH states are \emph{composite fermions} (CFs)~\cite{Jain89} -- electrons bound to vortices. Due to the nonperturbative nature of vortex attachment, predicting the measured STM signal becomes a highly nontrivial task, as anticipated in early theoretical works~\cite{Rezayi87a, He93b, Haussmann96}. At low electron densities, insufficient to form a FQH state, the LDOS can be analytically computed~\cite{MacDonald2010, LihKing2011, Chaudhary2019} and shown to consist of a series of peaks at energies given by the Haldane pseudopotentials~\cite{Haldane83}, providing a useful characterization of the relevant energy scales in a Landau level (LL)~\cite{Yoo19, Farahi2023}. 
By contrast, the understanding of the detailed structure of the STM spectra in the FQH regime has so far been lacking.       
For Jain states at electron filling factors $\nu{=}1/3, 2/5, 3/7$, etc., it was previously argued that a single hole and electron excitations have finite overlap on a tightly-bound state of multiple CFs, which should manifest as a resonance in LDOS~\cite{Jain05}. However, the recent high-resolution experiments on Bernal stacked bilayer graphene have reported \emph{multiple} sharp resonances for various FQH states realized in this system ~\cite{Hu23}. Moreover, the observed pattern of resonances displayed an intriguing asymmetry between the addition or removal of an electron.

In this paper, we show that LDOS, measured by scanning tunneling spectroscopy on a FQH state, consists of multiple peaks that can be naturally interpreted as CF ``Lambda levels'' ($\Lambda$Ls) -- the analogs of LLs of electrons~\cite{Jain07}, see Fig.~\ref{fig: sketch} for a schematic summary. We develop an efficient method for extracting LDOS spectra of FQH states belonging to the Jain sequence using CF wave functions, and we confirm its accuracy against exact diagonalization simulations. The interpretation of LDOS spectra in terms of $\Lambda$Ls explains the strong asymmetry in the numerically computed spectra for the addition vs. removal of an electron, which relates to the well-known asymmetry between CF quasiparticles and quasiholes. Implications for future experiments and potential uses of STM as a probe of new FQH states extending beyond the non-interacting CF paradigm are also discussed. 

\begin{figure*}
    \includegraphics[width=0.98\textwidth]{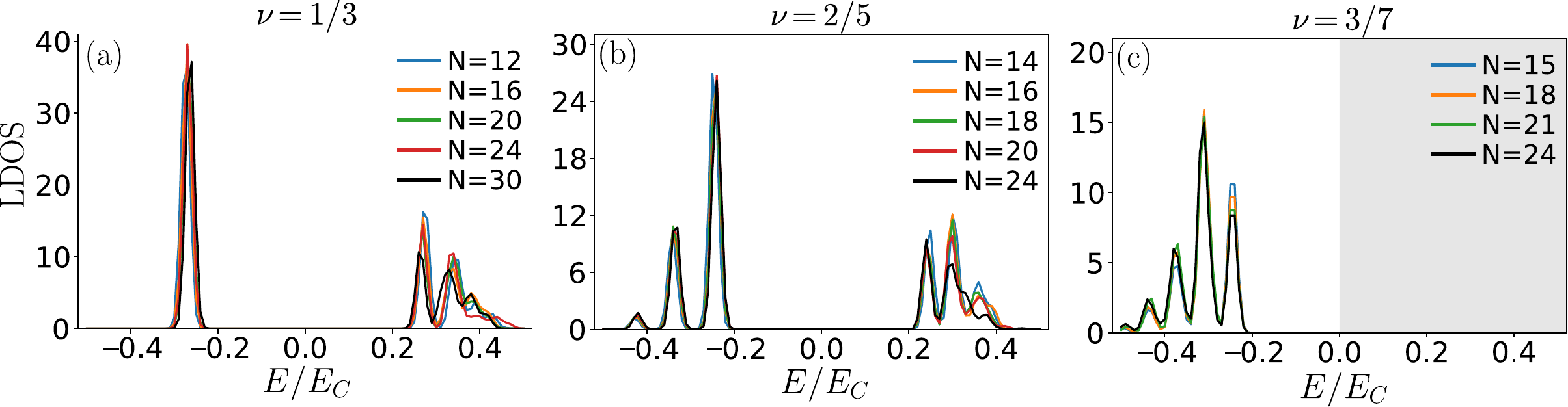}
    \caption{LDOS for FQH states at $\nu{=}1/3$, $2/5$, and $3/7$, obtained using CF theory, reveals distinctive peak structures on the hole and particle sides. We include the full CF space for the hole side and do a truncation $E_K^*{\leq} 2$ for the particle side. For the hole side, the data is well-converged across a range of systems shown in the legend. The computation of the particle side of LDOS is computationally intensive due to the unbounded growth of CF space, hence convergence was only achieved at $\nu{=}1/3, 2/5$, and thus the particle side has been omitted for $3/7$. All energies are quoted in units of $E_C{=}e^2/\varepsilon\ell$.
    }
    \label{fig: LDOS}
\end{figure*}

{\bf \em CF model for LDOS.---}We consider tunneling an electron into a FQH ground state $|\Omega\rangle$ with $N$ electrons. We assume the electrons are on the surface of a sphere, with a Dirac monopole at the center carrying $2Q$ flux quanta~\cite{Haldane83}. The radius of the sphere is $R{=}\sqrt{Q}\ell$, where $\ell{=}\sqrt{\hbar c/eB}$ is the magnetic length at magnetic field $B$. We consider Coulomb interaction for simplicity and express all energies in units of $E_C {\equiv} e^2/\varepsilon\ell$. We will focus on uniform FQH states residing in the lowest LL (LLL), with orbital angular momentum $L{=}0$. The electron is tunneled into the north pole, which is defined by the LL orbital with $L_{z}{=}Q$, and the corresponding energy-resolved LDOS is~\cite{Papic18, Pu22}
\begin{eqnarray}\label{eq: ldos}
\notag {\rm LDOS}(E, L_{z}{=}Q) &=& 
\sum_n \delta(E-E_n^-)|\langle n|c_{-Q}|\Omega\rangle|^2 \\
&+& \sum_n \delta(E-E_n^+)|\langle n|c_{Q}^\dagger|\Omega\rangle|^2,
\end{eqnarray}
where $c$ and $c^\dagger$ are the electron annihilation and creation operators, $n$ runs over all eigenstates with energy $E_n^\pm$, for $N{\pm}1$ electrons at the same flux $2Q$ as in the ground state.  For convenience, we will always assume that the tunneling process involves removing an electron from the south pole or adding it to the north pole, resulting in a state with $L{=}L_{z}{=}Q$. This can be assumed without any loss of generality since the FQH ground state is uniform. Henceforth, we will suppress the $L_{z}$ dependence. 

In principle, one could evaluate LDOS in Eq.~\eqref{eq: ldos} by brute force, obtaining all the eigenstates via exact diagonalization. However, this severely limits the accessible system sizes, amplifying the finite-size effects; furthermore, it sheds little light on the physics behind any of the observed LDOS features. We turn to CF theory~\cite{Jain89, Jain07} to overcome these obstacles. In CF theory~\cite{Jain89, Jain07}, Jain states at fillings $\nu{=}n/(2pn{\pm}1)$ are mapped into integer quantum Hall (IQH) states of CFs carrying $2p$ vortices and filling $n$ $\Lambda$Ls in a reduced effective flux of $2Q^{*}{=}2Q{-}2p(N{-}1)$ (all CF quantities are marked by a $*$ superscript). An example of $\nu{=}2/5$ is given in Fig.~\ref{fig: sketch}(b) with the shaded holes also filled. The Jain wave functions for these FQH states are given by $\Psi_{\nu{=}n/(2pn{\pm}1)}{=}\mathcal{P}_{\rm LLL}\Phi_{{\pm}n}\Phi_1^{2p}$, where $\Phi_n$ is the Slater determinant wave function of $n$ filled LLs ($\Phi_{{-}n}{=}[\Phi_{n}]^{*}$), and $\mathcal{P}_{\rm LLL}$ is the projection operator to LLL. From here on in, for convenience, we shall restrict our discussion to the $n/(2n{+}1)$ Jain states.

In CF theory, the removal of an electron from an FQH ground state (i.e., creating a hole excitation) is equivalent to creating $2n{+}1$ holes in the lowest $n$ $\Lambda$Ls ($n{=}0,1,2,{\cdots}$), as shown in Fig.~\ref{fig: sketch}(b). We refer to the ``CF space'' as the Hilbert space of $2n{+}1$ holes with $L{=}L_{z}{=}Q$ in the lowest $n$ $\Lambda$Ls. While the dimension of the Hilbert space of $N{-}1$ electrons in the LLL with $2Q$ fluxes and $L{=}L_{z}{=}Q$ grows exponentially with $N$, the dimension of the CF space is a constant (up to linear dependencies for small $N$): for example, at $\nu{=}1/3$ there is a unique state, at $\nu{=}2/5$ there are 3 states, at $\nu{=}3/7$ there are 27 states, etc.  This makes the diagonalizations in CF space much more efficient than in the full LLL space. Furthermore, CF theory provides explicit wave functions for the basis states that span the CF space, which we write as $\Psi_{\nu{=}n/(2n{+}1)}^{{-}, i}{=}\mathcal{P}_{\rm LLL}\Phi_n^{{-}, i}\Phi_1^{2}$. Here, $\Phi_n^{{-}, i}$ denote Slater determinants, indexed by $i$, of $N{-}1$ particles, which is equivalent to $2n{+}1$ holes in the lowest $n$ $\Lambda$Ls at flux $2Q^{*}{=}2Q{-}2(N{-}2)$ fluxes, with total angular momentum $L{=}L_{z}{=}Q$. These basis states, however, are not orthogonal and we perform CF diagonalization (CFD)~\cite{Mandal02} to obtain orthonormal states in the CF space, as detailed in the Supplemental Material (SM)~\cite{SOM}. 

Similarly, we consider adding one electron at the north pole. This is equivalent to creating $2n{+}1$ particles in the $m{\geq} n$ $\Lambda$Ls, as shown in Fig.~\ref{fig: sketch}(c). The CF space is now spanned by states of the form $\Psi_{\nu{=}n/(2n{+}1)}^{{+}, i}{=}\mathcal{P}_{\rm LLL}\Phi_n^{{+}, i}\Phi_1^{2}$, where $\Phi_n^{{+}, i}$ are Slater determinants for $(2n{+}1)$ particles in the $\Lambda$Ls with index $\geq n$ at flux $2Q^{*}{=}2Q{-}2N$ with angular momentum $L{=}L_{z}{=}Q$. Naively, one might think the dimension of the CF space for adding one electron would be infinite, as the $\Lambda$Ls do not have an upper bound. Nevertheless, this is not true because the CF space is still a subspace of the original Hilbert space of $N{+}1$ electrons in the LLL. This implies that the basis states of the CF space are not linearly independent. We classify the CF basis states by their effective kinetic energy, $E_K^*{=}\sum_{l{=}1}^{2n{+}1}(n_{l}{-}n){-}E_0$, where $n_{l}$ is $\Lambda$L index of the $l^{\rm th}$ particle and $E_0$ is the minimal value of $\sum_{l{=}1}^{2n{+}1}(n_{l}{-}n)$ in the $L_{z}{=}Q$ sector. For instance, $E_{0}{=}7$ for the configuration shown in Fig.~\ref{fig: sketch}(c). The CF kinetic energy is expressed in units of $\hbar\omega^{*}_{c}$, where $\omega^{*}_{c}$ is the CF cyclotron frequency. In practice, we carry out CFD in the CF subspace obtained by imposing an upper cutoff on $E_K^*$ and checking for convergence as this cutoff is increased. Similarly, $E_K^*$ is defined for the hole side with the replacement $n_l{-}n{\rightarrow} n{-}n_l$ [$E_{0}{=}1$ for the state shown in Fig.~\ref{fig: sketch}(b).]. 

We refer to the orthonormal states obtained from CFD as $|\alpha_n^{{\pm}}\rangle$ for adding or removing electrons, with the corresponding energy eigenvalues $E_n^{{\pm}}$, and define their corresponding overlaps with the electron and hole excitations as $\eta_n^{{+}}{=}\langle\alpha_n^{{+}}|c^\dagger_Q| \Omega \rangle$ and $\eta_n^{{-}}{=}\langle\alpha_n^{{-}}|c_{{-}Q}|\Omega\rangle$,  where $|\alpha_n^{\pm}\rangle$, $c^\dagger_Q|\Omega\rangle$ and $c_{{-}Q}|\Omega\rangle$ are all normalized to unity. Following Eq.~\eqref{eq: ldos}, we define the CF approximation to LDOS as 
\begin{equation}
{\rm LDOS}(E){=} \sum_n \left(\delta(E{-}E_n^{-})|\eta_n^{-}|^2 {+} \delta(E{-}E_n^{+})|\eta_n^{+}|^2 \right).
\label{eq: ldos_CF}
\end{equation}

When plotting the LDOS, we express the energy relative to the chemical potential~\cite{He93b}, including the appropriate electrostatic corrections as described in SM~\cite{SOM}. We smear the delta function in Eq.~\eqref{eq: ldos_CF} by a Gaussian of width $\sigma{=}0.01 E_C$ for easier visualization.

{\bf \em Results.---}We have benchmarked the accuracy of the above CF computation of LDOS against exact diagonalization, finding excellent agreement within the system sizes accessible to both methods~\cite{SOM}. Although CF theory is known to successfully capture the ground states and low-lying excitations of Jain states~\cite{Balram13, JainHalperin2020}, the accuracy of its LDOS approximation is nevertheless remarkable, given the smallness of the CF subspace and the high energies probed by electron tunneling into a FQH state. 

Our main results are presented in Fig.~\ref{fig: LDOS}, which shows the LDOS of FQH states at $\nu{=}1/3$, $2/5$, and $3/7$, obtained by the CF method. These plots reveal characteristic sequences of LDOS peaks, e.g., one peak on the hole side of $\nu{=}1/3$, three peaks on the particle side of $\nu{=}1/3$, and three peaks on both the hole and particle sides of $\nu{=}2/5$. These features are well-converged across a range of system sizes, thus they act as a universal ``fingerprint'' of each state, whose origin will be elucidated below. The LDOS is particularly well-converged on the hole side, where we kept \emph{all} basis states in the relevant CF space. On the particle side, the CF basis is unbounded and we must enforce an explicit truncation to keep the computation tractable. The numerics become computationally more intensive as the number of CF basis states increases. In Fig.~\ref{fig: LDOS}, we restricted the CF subspace to states with $E_K^*{=}0,1,2$, which was sufficient to converge the particle side of LDOS at $\nu{=}1/3$ and $2/5$. For $\nu{=}3/7$, however, this was insufficient to achieve convergence \cite{SOM}, and we do not show this data in Fig.~\ref{fig: LDOS}.

\begin{figure}[tb]
    \centering    
    \includegraphics[width=\linewidth]{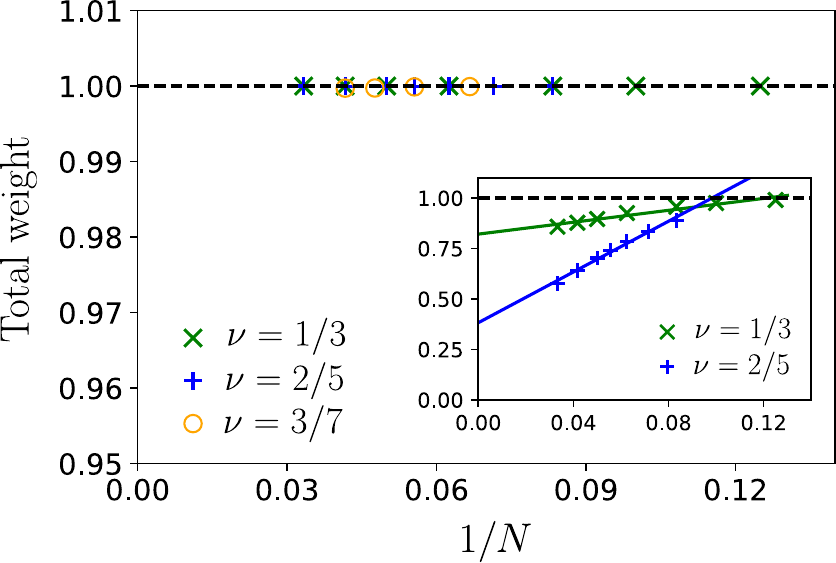}
    \caption{Total weight $\sum_n|\eta_n^{{-}}|^2$, which represents the amplitude of $c_{Q}|\Omega \rangle$ in the CF space, is unity. This shows the hole state is fully contained within the CF space. Inset: Total weight $\sum_n|\eta_n^{{+}}|^2$, i.e., the amplitude of the state $c^\dagger_{Q}|\Omega\rangle$ in the CF space. Lines are linear fits through the data. These amplitudes can be increased by increasing the value of $E_K^{*}$ at which we truncate (see text, we choose $E_K^*{\leq}2$). 
   }
    \label{fig: overlaps}
\end{figure}

To quantify the convergence of the CF calculation of LDOS, it is instructive to compute the total weight of states $c_{-Q}|\Omega\rangle$ and $c^\dagger_{Q}|\Omega\rangle$ in the CF space. If the total weight extrapolates to a  value of order unity in the thermodynamic limit, we can be confident that the CF approximation is capturing the key spectral features associated with the addition or removal of an electron from a FQH state. These weights are shown in Fig.~\ref{fig: overlaps} for $\nu{=}1/3$, $2/5$, and $3/7$ FQH states as a function of $1/N$. We see that the hole excitation fully resides within CF space, i.e., the total overlap of $c_{{-}Q}|\Omega\rangle$ on the CF basis is unity, and thus the hole side of the LDOS is fully captured by the CF theory. On the other hand, due to the imposed truncation, the support of $c^\dagger_{Q}|\Omega\rangle$ in the truncated CF space is generally less than 1, thus the particle side of LDOS is only partly captured by CF theory. In particular, in the thermodynamic limit, the truncated CF space captures about $80\%$ of the electron excitation for $\nu{=}1/3$ and $40\%$ for $\nu{=}2/5$. This number is expected to increase as we include more states with higher $E_K^*$.  

We note that Ref.~\cite{Jain05} computed overlaps similar to Fig.~\ref{fig: overlaps} by keeping only the unique state with the lowest $E_K^*$, for both the particle and hole sides. By contrast, we include the full CF space on the hole side and a much bigger portion of CF space on the particle side. Nevertheless, even restricting to the unique state with the lowest $E_K^*$, we find overlaps larger than those reported in Ref.~\cite{Jain05}. Our results have been additionally verified by direct calculation in the Fock space for small $N$~\cite{Sreejith13, Balram20a}. 

\begin{figure}[bt]
    \centering    
    \includegraphics[width=0.98\linewidth]{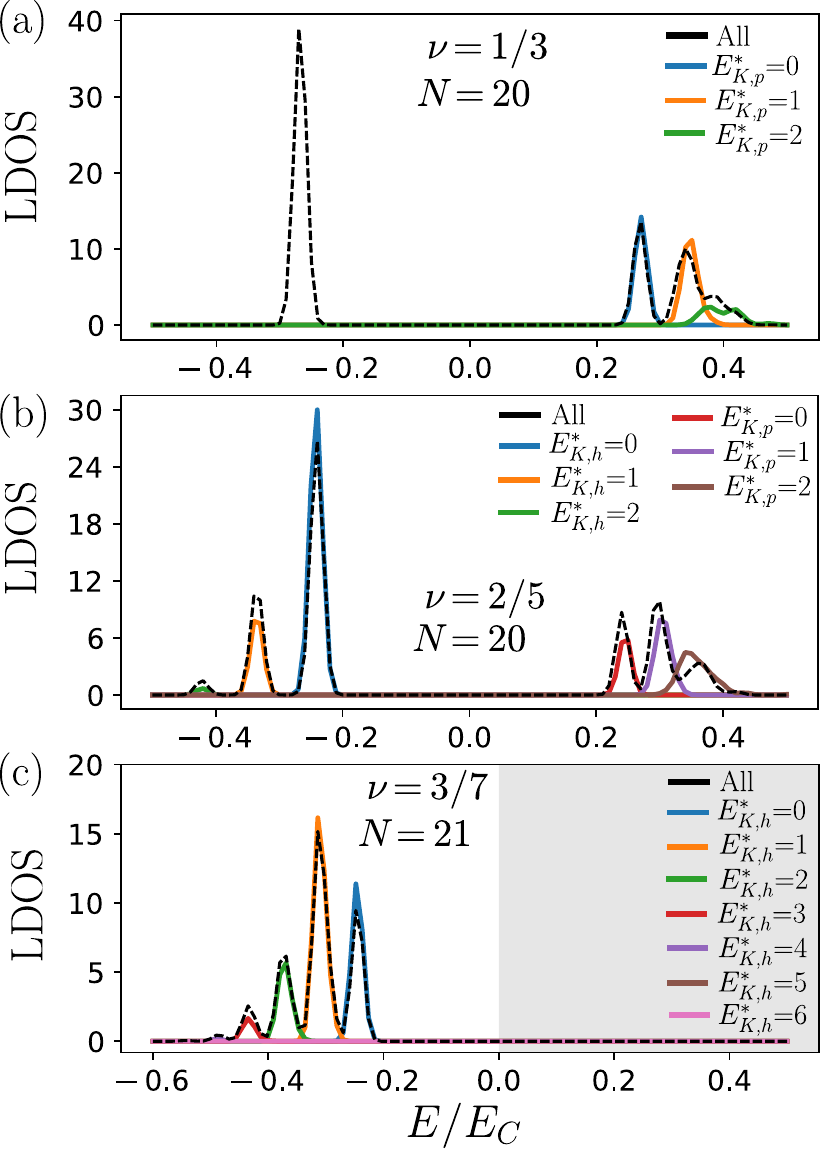}
    \caption{
    Reconstructing the full LDOS by populating successive $\Lambda$Ls with CFs in the $\nu{=}1/3, 2/5$ states with $N=20$ electrons (a) and  $\nu{=}3/7$ with $N{=}21$ electrons. (c) The plots show the LDOS in different CF subspaces made up of states with a single value of $E_K^*$ for the particle or hole sides, indicated in the legend, after removing all contributions with lower values of $E_K^{{*}}$. Each LDOS peak is associated with a definite CF kinetic energy. Black dashed lines represent the full LDOS from Fig.~\ref{fig: LDOS}.  
    } 
    \label{fig: LDOS_Lambda}
\end{figure}
{\bf \em Relating LDOS peaks to $\Lambda$Ls.}---The CF theory not only allows us to model LDOS quantitatively but also provides an understanding of the origins of peaks.  We surmise that each peak can be associated with CF states that carry a well-defined value of the kinetic energy, $E_K^*$. Numerical tests of this hypothesis are summarized in Fig.~\ref{fig: LDOS_Lambda}.

We first look at the simplest case: the hole side of $\nu{=}1/3$. In this case, there is only a unique state in the CF space with $L{=}L_{z}{=}Q$, and it is clear from Fig.~\ref{fig: LDOS} that this CF state accounts for the only peak seen in this case~\cite{Rezayi87a}.  Next, we consider the particle side of $\nu{=}1/3$. We first look at the CF space of the $E_K^*{=}0$ state, where there is only one state. We find that this state (blue line) perfectly captures the lowest energy peak on the particle side as shown in Fig.~\ref{fig: LDOS_Lambda}(a). Next, we go to the subspace made up of states with $E_K^*{=}1$. Unlike the IQH case, the states with different $E_K^*$ are not naturally orthogonal to each other~\cite{Dev92, Wu95, Mandal02, Balram13}. Therefore, we apply the Gram-Schmidt procedure to obtain a new basis for $E_K^*{=}1$ which is orthogonal to the states with $E_K^*{=}0$. We then carry out CFD on this new basis to obtain the LDOS~\cite{SOM}. After this, we find the states (orange line) in the resultant subspace perfectly capture the second peak on the particle side. Finally, we repeat the above procedure for $E_K^*{=}2$ (green line). While the peak height corresponding to this subspace does not perfectly match the third peak, the energy range where it occurs agrees well. We have performed a similar analysis for $\nu{=}2/5$ and the hole side of $\nu{=}3/7$, finding similarly good agreement, see Fig.~\ref{fig: LDOS_Lambda}(b)-(c).

{\bf \em Conclusions and discussion.---}We have proposed that STM experiments, performed on FQH states belonging to the primary Jain sequence $\nu{=}n/(2n{+}1)$, can reveal sharp resonances associated with CF-LLs. For sufficiently small $n{\leq} 3$, these features were shown to be robust across various system sizes, with good agreement between CF theory and exact numerics, despite the high energy of excitations probed in this setup.

A recent experiment on Bernal stacked bilayer graphene~\cite{Hu23} has reported three LDOS peaks for electron-like excitation for $\nu{=}3/5$, which would correspond to our predicted $\nu{=}2/5$ hole-like excitation. In contrast, in the same experiment, $\nu{=}2/3$ appears to show not just one electron-like peak but two, which our model would predict to have just one resonance. Future experiments would be required to make contact with our theoretical prediction. In particular, an important test for identifying the predicted $\Lambda$L peaks would be precise measurements of their field dependence. All the peaks in this work are interaction-driven, hence their positions should scale as $E_C{ \propto} \sqrt{B}$. 
This would distinguish them, e.g., from additional features associated with the spin physics of CFs. Furthermore, for a direct comparison with experiment, it would be necessary to also include the effects of disorder and screening. Nevertheless, while these effects may impact the precise shape of LDOS spectra, we expect the correspondence between peaks and CF kinetic energy to continue to hold. We reserve the detailed investigation of these effects for future work.

For larger $n$, i.e., filling factors approaching the $\nu{=}1/2$, the calculations presented above are expected to become significantly harder due to the faster growth of CF space. Moreover, the CF cyclotron energy becomes smaller as the CF Fermi liquid state at $1/2$ is approached, implying that the identification of LDOS peaks with CF $\Lambda$Ls may no longer be as straightforward as in Fig.~\ref{fig: LDOS_Lambda} above. Due to the ``aliasing'' problem on the sphere~\cite{Ambrumenil89}, the study of such FQH states would also require larger system sizes, presenting an interesting challenge for theory.    

Finally, our study opens up several interesting questions. Additional degrees of freedom such as spin, layer, valley, etc. will naturally give rise to richer LDOS spectra that could be studied using the present method. On the other hand, for higher-order CF states such as those at $\nu{=}2/7,~2/9$ (where non-CF partonic features might be present~\cite{Balram21d}) and for states beyond the non-interacting CF paradigm, e.g., the paired $\nu{=}5/2$ state, a different approach altogether may be needed to understand the LDOS. Consequently, we expect some violations of the peak structure identified above, which could serve as a diagnostic tool for unconventional physics extending beyond the standard CF theory at high energies~\cite{Balram21d, Nguyen22}. 

{\sl Noted Added: }During the completion of this work, we became aware of a related work by Gattu \emph{et al.}~\cite{Gattu2023}, which also studies the CF description of LDOS of FQH states. 

\begin{acknowledgments}
	
{\bf \em Acknowledgments.---}We thank Ying-Hai Wu for useful discussion. We acknowledge support by the Leverhulme Trust Research Leadership Award RL-2019-015 and Royal Society International Exchanges Grant IES$\backslash$R2$\backslash$202052. 
This research was supported in part by grants NSF PHY-1748958 and PHY-2309135 to the Kavli Institute for Theoretical Physics (KITP). N.R. acknowledges support from the QuantERA II Programme which has received funding from the European Union’s Horizon 2020 research and innovation programme under Grant Agreement No 101017733. Computational portions of this research work were carried out on ARC3 and ARC4, part of the High-Performance Computing facilities at the University of Leeds, UK, and the Nandadevi supercomputer, which is maintained and supported by the Institute of Mathematical Science's High-Performance Computing Center. Some of the numerical calculations were performed using the DiagHam package~\cite{DiagHam}. ACB thanks the Science and Engineering Research Board (SERB) of the Department of Science and Technology (DST) for funding support via the Mathematical Research Impact Centric Support (MATRICS) Grant No. MTR/2023/000002. AY, YH, Y-CT, and MH acknowledge primary support from  DOE-BES grant DE-FG02-07ER46419 and the Gordon and Betty Moore Foundation’s EPiQS initiative grant GBMF9469. Other support to AY from NSF-MRSEC through the Princeton Center for Complex Materials NSF-DMR- 2011750, ARO MURI (W911NF-21-2-0147), and ONR N00012-21-1-2592 are also acknowledged. 
\end{acknowledgments}

$^*$These authors have contributed equally to this work.

\newpage 
\cleardoublepage

\onecolumngrid
\begin{center}
\textbf{\large Supplemental Online Material for ``Fingerprints of Composite Fermion Lambda Levels in Scanning Tunneling Microscopy" }\\[5pt]
\begin{center}
 {\small Songyang Pu$^{1,2*}$\orcidlink{https://orcid.org/0000-0003-0144-1548}, Ajit C. Balram$^{3,4*}$\orcidlink{0000-0002-8087-6015}, Yuwen Hu$^{5}$, Yen-Chen Tsui$^{5}$, Minhao He$^{5}$, Nicolas Regnault$^{6,5}$, Michael P. Zaletel$^{7}$, Ali Yazdani$^{5}$, Zlatko Papi\'c$^{1}$ }
\end{center}

\begin{center}
{\sl \footnotesize $^{1}$School of Physics and Astronomy, University of Leeds, Leeds LS2 9JT, United Kingdom

$^{2}$Department of Physics and Astronomy, The University of Tennessee, Knoxville, TN 37996, USA

$^{3}$Institute of Mathematical Sciences, CIT Campus, Chennai, 600113, India

$^{4}$Homi Bhabha National Institute, Training School Complex, Anushaktinagar, Mumbai 400094, India

$^{5}$Department of Physics, Princeton University, Princeton, NJ 08544, USA

$^{6}$Laboratoire de Physique de l’Ecole normale sup\'erieure,ENS, Universit\'e PSL, CNRS, Sorbonne Universit\'e

$^{7}$Department of Physics, University of California at Berkeley, Berkeley, CA 94720, USA
}
\end{center}
\vspace{0.1cm}
\begin{quote}
{\small In this Supplemental Material, we present details of CF calculations including (i) the description of the wave functions for the ground state and excited states containing an electron or hole on one of the poles of the sphere; (ii) the CF diagonalization procedure and computation of LDOS; (iii) comparison of CF basis dimensions for several filling factors on the particle and hole side; (iv) a simple toy model for the amplitude of LDOS peaks; (v) a discussion of convergence of CF results and their comparison against LDOS obtained in exact numerics; (vi) discussion of the LDOS spectrum on the particle side of $\nu=3/7$.
}\\[20pt]
\end{quote}
\end{center}

\setcounter{equation}{0}
\setcounter{figure}{0}
\setcounter{table}{0}
\setcounter{page}{1}
\setcounter{section}{0}
\makeatletter
\renewcommand{\theequation}{S\arabic{equation}}
\renewcommand{\thefigure}{S\arabic{figure}}
\renewcommand{\thesection}{S\Roman{section}}
\renewcommand{\thepage}{\arabic{page}}
\renewcommand{\thetable}{S\arabic{table}}

\vspace{0cm}

\section{Composite fermion wave functions}

In this section, we briefly review the LLL-projected CF wave functions following the standard method outlined in Ref.~\cite{Jain07}. In the spherical geometry, it is convenient to work in spinor coordinates~\cite{Haldane83}
\begin{equation}
    u=\cos(\theta/2)e^{i\phi/2}, \quad 
    v=\sin(\theta/2)e^{-i\phi/2},    
\end{equation}
where $\theta$ and $\phi$ are the polar and azimuthal angles on the sphere. The orbitals in the LLL are monopole harmonics~\cite{Wu76}: 
\begin{equation}
\label{orb}
Y_{QQm}=\left[{2Q+1\over 4\pi}\binom{2Q}{Q-m}\right]^{1/2}(-1)^{Q-m}v^{Q-m}u^{Q+m}.
\end{equation}
The CF wave functions for ground states at filling factors $\nu{=}n/(2pn{+}1)$ are given by
\begin{equation}
\label{wf}
    \Psi^{i}_{n/(2pn+1)}=\mathcal{P}_{\rm LLL}\Phi_n^{i}\Phi_1^{2p}, \quad \Phi_1=\prod_{i<j}\left(u_iv_j-u_jv_i\right),
\end{equation}
where $\mathcal{P}_{\rm LLL}$ is the LLL projection operator and  $\Phi_n^{i}$ is the Slater determinant of all electron coordinates and occupied $\Lambda$L orbitals. If $\Phi_n^{i}$ has only the lowest $n$ $\Lambda$Ls occupied, Eq.~\eqref{wf} describes the Jain state at $\nu{=}n/(2pn{+}1)$~\cite{Jain89}.

To evaluate the wave functions, we implement the Jain-Kamilla projection technique \cite{Jain97, Jain97b}. The projected CF wave function is 
\begin{equation}
\label{project}
    \Psi_{n/(2pn+1)}={\rm Det}[\phi_{n,m,j}[u_i,v_i]]\prod_{i<j}\left(u_iv_j-u_jv_i\right)^{2p},
\end{equation}
where $\phi_{n,m,j}$ is the projected ``orbital" with $n$ the $\Lambda$L index, $m$ the $L_{z}$ index, and $j$ the coordinate index. The explicit expression is given by
\begin{equation}
    \phi_{n,m,j}=N_{q,n,m}(-1)^{q+n-m}{(2Q+1)!\over(2Q+n+1)!}u_j^{q+m}v_j^{q-m}\sum_{s=0}^n(-1)^s\binom{n}{s}\binom{2q+n}{q+n-m-s}u_j^sv_j^{n-s}P_j[s,n-s],
\end{equation}
where  $Q$ is the physical flux number generating LLs, $q{=}Q{-}2p(N{-}1)$ is the effective flux number generating $\Lambda$Ls, and $N_{q,n,m}{=}\left({2q{+}2n{+}1\over 4\pi}{(q{+}n{-}m)!(q{+}n{+}m)!\over n!(2q{+}n)!}\right)^{1/2}$. The function $P_j[s,n{-}s]$ is evaluated recursively by the following relations
\begin{gather*}
    P_j[s,n-s]\equiv\left(pf_j(1,0)+{\partial\over\partial u_j}\right)^s\left(pf_j(0,1)+{\partial\over\partial v_j}\right)^s\mathds{1},\\
    f_j(\alpha,\beta)\equiv \sum_{k,k\neq j}\left({v_k\over u_jv_k-u_kv_j}\right)^\alpha \left({-u_k\over u_jv_k-u_kv_j}\right)^\beta,\\
    {\partial\over \partial u_j}f_j(\alpha,\beta)=-(\alpha+\beta)f_j(\alpha+1,\beta),\\
    {\partial\over \partial v_j}f_j(\alpha,\beta)=-(\alpha+\beta)f_j(\alpha,\beta+1).
\end{gather*}

\section{Hole and Particle}

In the section, we present the expressions for the wave functions of $c_{{-}Q}|\Psi_{n/(2pn{+}1)}\rangle$ and $c^\dagger_{Q}|\Psi_{n/(2pn{+}1)}\rangle$~\cite{Jain05}. We first note that acting with $c^\dagger_{Q}$ is equivalent to putting an electron at the north pole and acting with $c_{{-}Q}$ is equivalent to removing an electron from the south pole. The hole state $c_{{-}Q}|\Psi_{n/(2pn{+}1)}\rangle$ can thus be viewed as a ground state with $N$ electrons with the first electron fixed at the south pole. Therefore, its wave function is the same as Eq.~\eqref{project} with the lowest $n$ $\Lambda$Ls fully filled and $u_1{=}0,v_1{=}e^{{-}i\phi/2}$.

It is somewhat more tedious to evaluate $c^\dagger_{Q}|\Psi_{n/(2pn{+}1)}\rangle$, which is essentially the anti-symmetrization of $Y_{QQQ}$ in Eq.~\eqref{orb} and Eq.~\eqref{project}. The final form is
\begin{equation}
\Psi_{c^\dagger_{Q}|\Psi_{n/(2pn+1)}\rangle}=\sum_{s=1}^{N+1}(-1)^{s(N-1)}Y_{QQQ}(u[s],v[s])\tilde{\Psi}^{\rm GS}_{n/(2pn+1)},
\end{equation}
where $\tilde{\Psi}^{\rm GS}_{n/(2pn+1)}$ is the Jain state given by Eq.~\eqref{project} with $N$ electrons $i{=}1,2,{\cdots}, s{-}1,s{+}1,{\cdots}, N{+}1$ fully filling the lowest $n$ $\Lambda$Ls.

\section{CF diagonalization}
\label{CFD}

We refer to the basis for the CF space discussed in the main text as $|\beta_n\rangle$. This space is spanned by wave functions of the form $\mathcal{P}_{\rm LLL}\Phi_n^{\pm}\Phi_1^{2p}$ for Jain states at $\nu{=}n/(2pn{+}1)$, where $\Phi_n^{{\pm}}$ are Slater determinants of particle (hole) states of integer quantum Hall states. In the case of removing one electron, the basis is finite and complete, while in the case of adding one electron, we do a truncation on the CF kinetic energy $E_K^*$ (see main text), which makes the basis incomplete. As shown in the main text, keeping states with $E_K^*{=}0,1,2$ captures the lowest three peaks of LDOS on the particle side to a satisfactory degree while missing some high-energy parts.

In general, the CF basis states are neither orthogonal nor linearly independent. To produce a set of orthogonal and linearly independent basis states from the CF states, we take three steps \cite{Mandal02, Balram13}. First, we calculate the overlap matrix $O_{m, n}{=}\langle \beta_m|\beta_n\rangle$ and Hamiltonian matrix $H_{m, n}{=}\langle \beta_m|V|\beta_n\rangle$, where $V(r)=1/r$ is the Coulomb interaction. 

Second, we do a singular value decomposition on the overlap matrix to obtain a linearly independent basis $|\tilde{\beta}_n\rangle$ and calculate $\tilde{O}_{m, n}{=}\langle \tilde{\beta}_m|\tilde{\beta}_n\rangle$ and $\tilde{H}_{m, n}{=}\langle \tilde{\beta}_m|V|\tilde{\beta}_n\rangle$. Note that the vectors $|\tilde{\beta}_n\rangle$ are linearly independent but are not necessarily orthogonal to each other. 

Third, the orthogonal eigenvectors $|\tilde{\alpha}_n\rangle$ of the interaction $V$ can be obtained by diagonalizing $\tilde{O}^{-1}\tilde{H}$ \cite{Balram15}. Finally, this gives us the eigenvalues $\epsilon_n$ and ortho-normalized eigenstates $|\alpha_n\rangle={|\tilde{\alpha}_n\rangle\over\sqrt{\langle\tilde{\alpha}_n|\tilde{\alpha}_n\rangle}}$.

\section{Dimension of CF space and truncations on the particle side}

As mentioned in the main text, a major advantage of our approach is that the dimension of the CF space is much smaller than the full Fock space of electrons in the LLL. The hole side fully resides within the CF space while a large portion of the particle side can be covered by keeping CF states up to $E_K^*{=}0,1,2$. While both $L^2$ and $L_{z}$ are good quantum numbers in our calculation, we carry out the composite fermion diagonalization (CFD) in the $L_{z}$ sector as the natural CF basis vectors are only eigenstates of $L_{z}$ (one can form $L^{2}$ eigenstates from these $L_{z}$ eigenstates using Clebsch-Gordan coefficients but that does not bring down the complexity of the calculation). In Table~\ref{Dim} below, we summarized the dimensions of the CF basis for all system sizes presented in this work. The actual dimensions of the CF space considered are equal to or smaller than these numbers, as $L^2$ has not been resolved. For instance, there are four states in the $L_{z}{=}Q$ sector for $\nu{=}2/5$ hole side while only three of them have $L{=}Q$. However, this does not change the LDOS results as states with $L^2{\neq} Q(Q{+}1)$ have zero overlaps with $c^\dagger_{Q}|\Psi_{n/(2pn{+}1)}\rangle$ or $c_{{-}Q}|\Psi_{n/(2pn{+}1)}\rangle$.
\begin{table*}
\makebox[\textwidth]{\begin{tabular}{|c|c|c|c|c|}
\hline
particle/hole&$\nu$&N&$D_1$&$D_2$\\ \hline
\multirow{3}{*}{hole}&1/3&$\geq 10$&1&1\\ \cline{2-5}
&2/5&$\geq 12$&4&4\\ \cline{2-5}
&3/7&$\geq 15$&39&39\\ \hline
\multirow{4}{*}{particle}&1/3&$\geq 10$&18&18\\ \cline{2-5}
&\multirow{3}{*}{2/5}&12&61&59\\ \cline{3-5}
&&14&61&60 \\ \cline{3-5}
&& $\geq 16$ &61&61\\ \hline
\end{tabular}}
\caption{The dimension of the CF basis in $L_{z}{=}Q$ sector before ($D_1$) and after ($D_2$) singular value decomposition. For simplicity, the $L^2$ is not resolved in our calculation. The CF basis fully covers the hole side and covers the $E_K^*{=}0,1,2$ on the particle side.\label{Dim}}
\end{table*}

The total weight captured in the CF space for the particle side can be increased by extending the truncation to higher $E_K^*$ since more CF states would then be included. In Table~\ref{Tru}, we illustrate the convergence of the weight $\sum_n|\eta_n^{{+}}|^2$ for different truncations at $\nu{=}1/3, 2/5$.
\begin{table*}
\makebox[\textwidth]{\begin{tabular}{|c|c|c|c|}
\hline
$\nu$&$E_K^*=0$&$E_K^*\leq 1$&$E_K^*\leq 2$\\ \hline
1/3&0.356&0.726&0.896\\ \hline
2/5&0.160&0.427&0.702\\ \hline
\end{tabular}}
\caption{The total weight captured in the CF space for particle side $\sum_n|\eta_n^{{+}}|^2$ for different truncations on $E_K^*$ for $N{=}20$ electrons at 1/3 and 2/5.
\label{Tru}
}
\end{table*}

\section{Evaluation of LDOS}

We refer to the orthonormal states obtained from CFD as $|\alpha_n^{{-}}\rangle$ ($|\alpha_n^{{+}}\rangle$) for the removal (addition) of an electron and the corresponding eigenvalues $\epsilon_n^{{-}}$ ($\epsilon_n^+$), where $n{=}0,1,2\cdots$ and the values are ordered by their absolute value. Defining the overlaps $\eta_n^{{+}}{=}\langle\alpha_n^{{+}}|c^\dagger_Q| \Omega \rangle$ and $\eta_n^{{-}}{=}\langle\alpha_n^{{-}}|c_{{-}Q}|\Omega\rangle$, the LDOS can be evaluated as 
\begin{equation}
    \mathrm{LDOS}(E)=\sum_{n,\gamma=\pm} {1\over\sqrt{2\pi}\sigma}{\rm exp}\left[-{(E-E_n^\gamma)^2\over 2\sigma^2}\right]|\eta^\gamma_n|^2.
\end{equation}
The eigenenergies are given by
\begin{equation}\label{eq: eigenenergies}
    E_n^{\pm}=\pm\left[\epsilon_n^{\pm}-{(N\pm 1)^2-1\over 2\sqrt{Q}}-\left(E_G-{N^2\over 2\sqrt{Q}}\right)\mp \mu\right],
\end{equation}
with $E_G$ being the energy of the Jain ground state and we have expressed the delta function as a Gaussian of width $\sigma$ for easier visualization of the spectral features (unless specified otherwise, we use $\sigma{=}0.01 E_C$). Furthermore, the energies are measured relative to the chemical potential~\cite{He93b}, 
\begin{equation}\label{eq: mu}
 \mu={1\over 2}\left[ \left( \epsilon_0^+-{(N+1)^2-1\over2\sqrt{Q}}\right) - \left(\epsilon_0^--{(N-1)^2-1\over2\sqrt{Q}}\right)\right]. 
\end{equation}
In both Eqs.~\eqref{eq: eigenenergies} and \eqref{eq: mu}, we have included the electrostatic corrections due to the positively charged background~\cite{Morf02}. 

In our calculation for $\nu{=}1/3,2/5$, $\epsilon_0^{{\pm}}$ in Eq.~\eqref{eq: mu} are the lowest energies obtained from CFD on the particle and hole sides. At $\nu{=}3/7$, however, we cannot get converged results for the particle side with only $E_K^*{=}0,1,2$. We decide to fix the chemical potential, or equivalently $\epsilon_0^+$ by estimating $E_0^+{-}E_0^-$. If the pairs of CF quasiparticles and quasiholes are non-interacting, this gap should be approximated by $2n{+}1$ times transport gaps $\Delta$ at $\nu{=}n/(2n{+}1)$ as we have $2n{+}1$ pairs of CF quasiparticle and quasihole. However, the constraint of the total angular momentum effectively brings those pairs close and the interaction between the quasiparticles and quasiholes is important to account for. This is verified in our results of $\nu{=}1/3, 2/5$, as $E_0^+{-}E_0^-$ is approximately $5\Delta$ for $1/3$ and $8\Delta$ for $2/5$. Therefore, we assume $E_0^{{+}}{-}E_0^{{-}}{=}14\Delta$, where the coefficient $14$ is chosen by estimation and $\Delta$ takes its \emph{thermodynamic limit} $0.035E_C$~\cite{Balram16d, Zhao22}. A different choice of the coefficient only shifts the LDOS spectra along the energy axis keeping the line shapes invariant. Based on this assumption, we have $\mu{=}7\Delta{+}\left(E_G{-}{N^2\over 2\sqrt{Q}}\right){-}\left(\epsilon_0^{{-}}{-}{(N{-}1)^2{-}1\over 2\sqrt{Q}}\right)$.

To alleviate the effect of finite sizes all energies shown in Fig. 2 and Fig. 4 in the main text are multiplied by a density correction factor $\sqrt{2Q\nu\over N}$~\cite{Morf86}.

\section{Toy model for the LDOS peaks' amplitude}

As we mentioned in the main text, the peaks in the LDOS spectrum have their origins in CF states with different effective kinetic energies. One may wonder whether it is possible to predict the peaks' amplitude intuitively without detailed calculations. The energy spacing between neighboring peaks should be $\hbar \omega_c^*$, and the amplitude of the $n$th peak is expected to be proportional to $d_n$ -- the number of independent CF states with $L{=}L_z{=}Q$ and $E_K^*{=}n$, times a factor $e^{{-}\lambda n}$ as the contribution from individual state is supposed to decay with energy. In short, the ansatz for the LDOS spectrum can take the form 
\begin{equation}
\label{eq: LDOS_naive}
    {\rm LDOS}(E)=\sum_n\delta(E-E_0-n\hbar\omega_c^*)d_ne^{-\lambda n}
\end{equation}
where $E_0$,  $\hbar\omega_c^*$, and $\lambda$ are treated as fitting parameters. 

\begin{figure}[b]
    \centering    
    \includegraphics[width=0.45\linewidth]{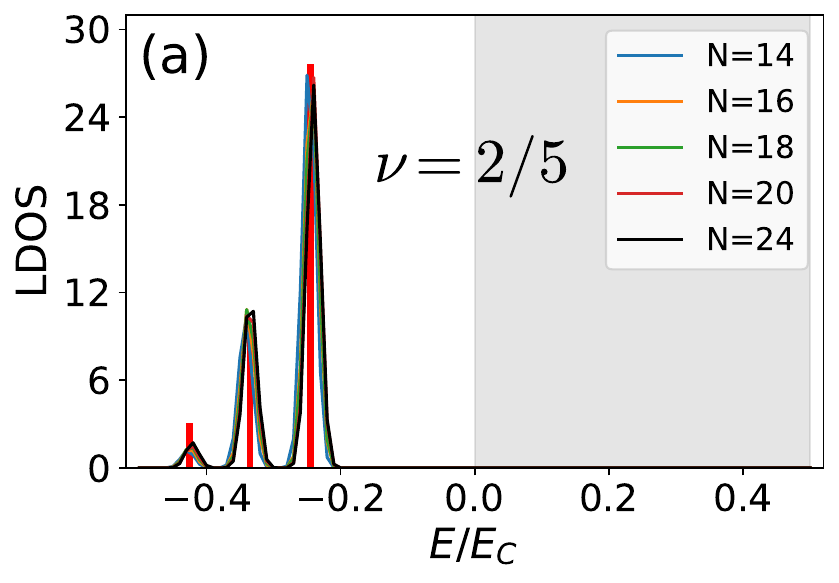}
     \includegraphics[width=0.45\linewidth]{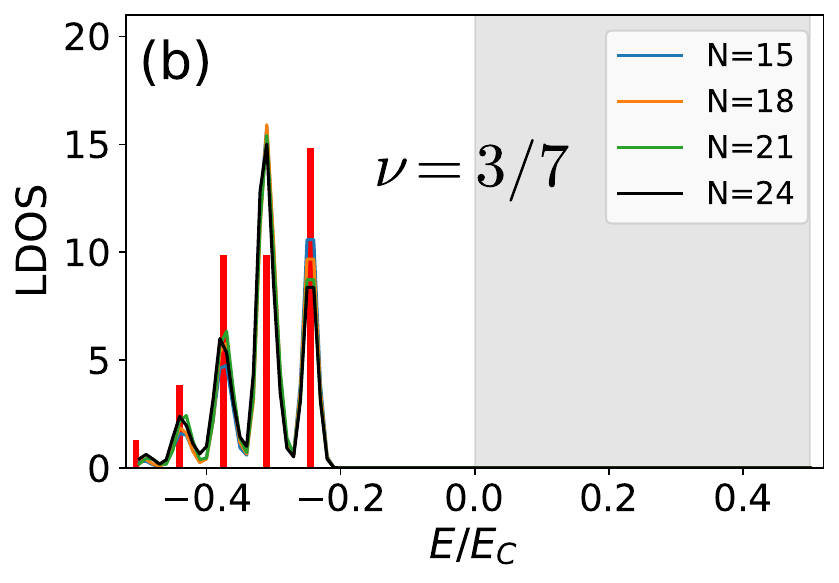}
    \caption{
    Fitting of the CF results presented in the main text by Eq.~\eqref{eq: LDOS_naive} with parameter values given in Table~\ref{fitting}. The smooth curves are reproductions of the results in the main text, while the red bars are fitted results according to Eq.~\eqref{eq: LDOS_naive}.
    }
    \label{fit}
\end{figure}

We tested whether Eq.~\eqref{eq: LDOS_naive} could be fitted to our quantitative CF results for $\nu{=}2/5$ and $3/7$ hole side. The fitting parameters along with $d_n$ are given in Table.~\ref{fitting}, while the fitting results are shown in Fig.~\ref{fit}. While the fitting is found to work fairly well for $\nu{=}2/5$, the fitting only captures the energy of the peaks for $\nu{=}3/7$. The disagreement in the amplitudes is expected, as the CF states with different $E_K^*$ are not orthogonal to each other. Therefore, it is not sufficient to only know the number of independent states $d_n$ at $E_K^*{=}n$, but also to remove the contribution from $E_K^*{<}n$, as we did in the main text and further explained in the following sections. Therefore, the simple model of Eq.~\eqref{eq: LDOS_naive} is not able to fully capture the LDOS.

\begin{table*}
\makebox[\textwidth]{\begin{tabular}{|c|c|c|c|c|c|}
\hline
$\nu$&number of $E_K^*$&$d_n$&$E_0$&$\hbar\omega_c^*$&$\lambda$\\ \hline
2/5&3&(1,1,1)&-0.245&0.09&1.1\\ \hline
3/7&7&(1,2,6,7,7,3,1)&-0.245&0.065&1.1\\ \hline
\end{tabular}}
\caption{Values of parameters in Eq.~\eqref{eq: LDOS_naive} to fit the CF results in Fig.~\ref{fit}.
\label{fitting}
}
\end{table*}

\section{Division of CF space by values of $E_K^*$}

In this section, we explain how we divide the CF space into subspaces based on values of $E_K^*$. First, there is only one state $|\alpha_0\rangle$ with $E_K^*{=}0$, and as it is mentioned in the main text, this state results in the lowest energy peak seen in LDOS.

Second, there is a set of CF states with $E_K^*{=}1$, and we label them as $|\beta_{1, i}\rangle$. To make them orthogonal to $|\alpha_0\rangle$, we implement the Gram-Schmidt procedure and obtain the new basis $|\bar{\beta}_{1, i}\rangle{=}\mathcal{N}\left(|\beta_{1, i}\rangle{-}| \alpha_0\rangle\langle\beta_{1, i}|\alpha_0\rangle\right)$, where $\mathcal{N}$ is the normalization factor. We then use this new basis for the CFD procedure mentioned above to obtain $|\alpha_{1, i}\rangle$.

The above procedure can be repeated to obtain the CF subspace with higher $E_K^*$. Suppose we have the basis $|\alpha_{m,i}\rangle$ for $m{<}n$ which satisfies $\langle\alpha_{m,i}|\alpha_{l,j}\rangle {=} \delta_{m,l}\delta_{i,j}$. To obtain $|\alpha_{n, i}\rangle$, we apply the CFD procedure to $|\bar{\beta}_{n, i}\rangle{=}\mathcal{N}\left(|\beta_{n, i}\rangle{-}\sum_{m{=}0}^{n{-}1}\sum_{j}| \alpha_{m,j}\rangle\langle\beta_{n, i}|\alpha_{m,j}\rangle\right)$, where $|\beta_{n, i}\rangle$ are the ``natural" CF states of the form of $\mathcal{P}_{\rm LLL}\Phi_n\Phi_1^{2p}$.

\section{Comparison between CF theory and exact diagonalization}

\begin{figure}[t]
    \centering    
    \includegraphics[width=0.32\linewidth]{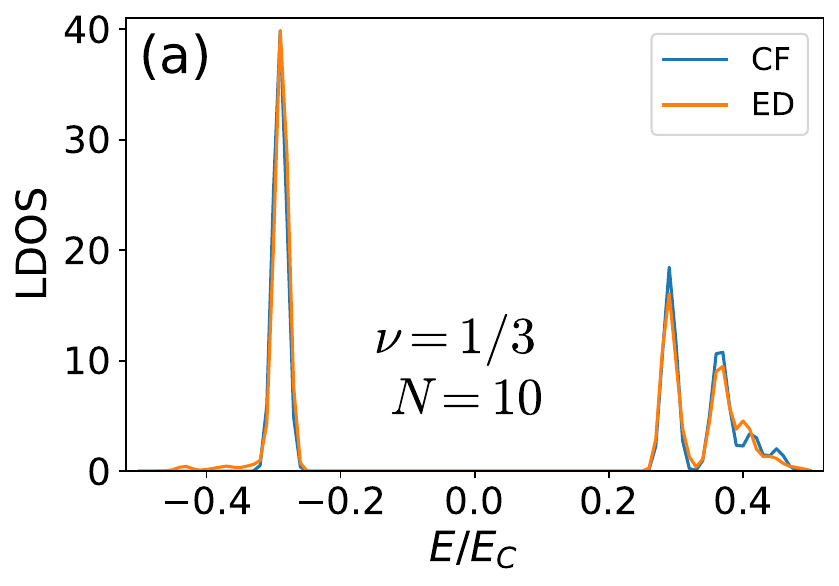}
    \includegraphics[width=0.32\linewidth]{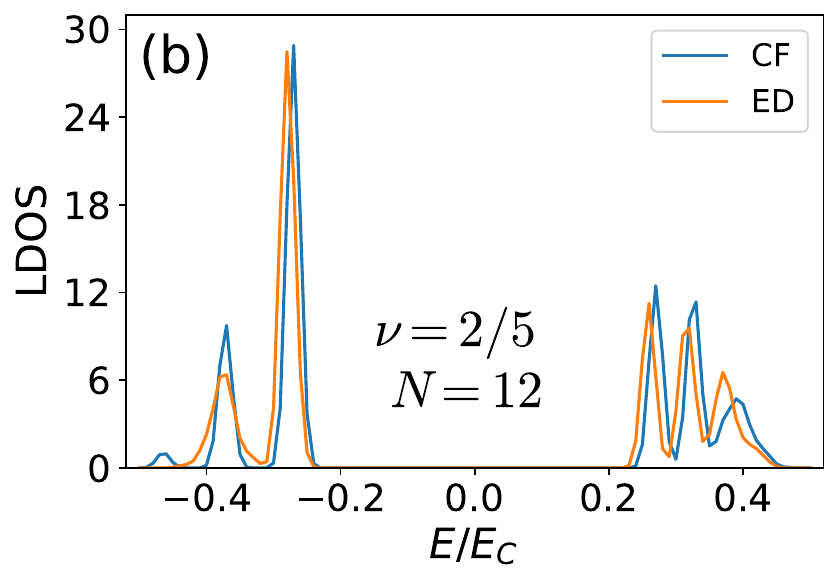}
     \includegraphics[width=0.32\linewidth]{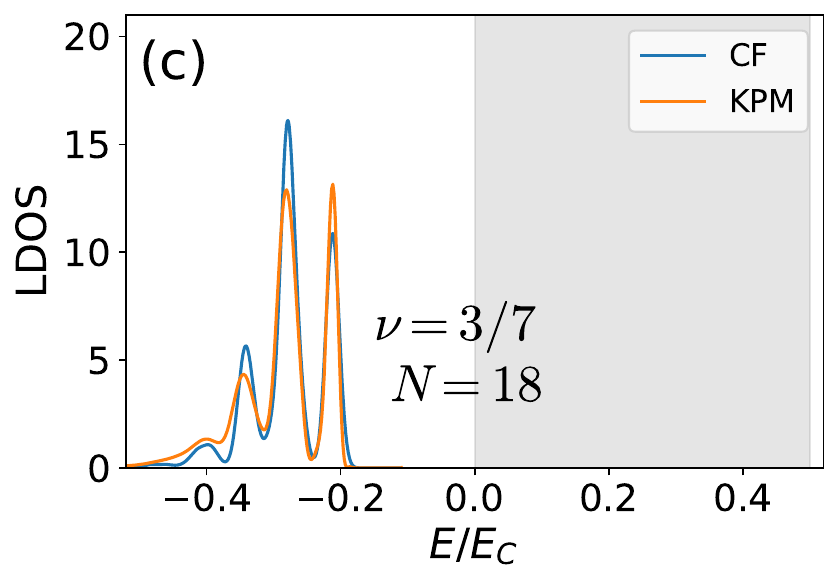}
    \caption{
    A comparison of the LDOS calculated by ED and CF methods for (a) $\nu{=}1/3$, $N{=}10$, (b) $\nu{=}2/5$, $N{=}12$, and (c) $\nu{=}3/7$, $N{=}18$. The CF space fully covers the hole side such that the squared overlap adds up to one. On the particle side, we keep the states with $E_K^*{=}0,1,2$ for both $\nu{=}1/3,2/5$.
    }
    \label{fig: ED}
\end{figure}

To confirm that the CF theory captures the main features of LDOS, we first compare the results with those obtained from ED in system sizes accessible to the latter. Fig.~\ref{fig: ED} confirms the quantitative agreement between CF and ED results for $\nu{=}1/3$ and $\nu{=}2/5$ states. The level of agreement is high, especially considering that we are probing high-energy excitations above the FQH ground state. While the CF space is much smaller than the full Hilbert space and a big portion of the high-energy part is missing in the CF space, the dominant peaks of LDOS are nearly fully captured in CF calculations, justifying the use of this method to compute LDOS. 

For $\nu{=}3/7$, the largest system size available to exact diagonalization, $N{=}15$, suffers from the ``aliasing'' issue~\cite{Ambrumenil89}. Consequently, this system size is not well representative of the intrinsic physics of the $\nu{=}3/7$ state in the thermodynamic limit. The hole side of the $N{=}15$ system aliases with 2 CF particles at $\nu{=}2/5$, while the particle side aliases with the $4/9$ Jain state. Hence, we looked at the next system size, $N{=}18$, with a corresponding Hilbert space dimension close to $400$ million, where ED can only access a few low-lying states. In this case, we compare our CF results against an iterative calculation of LDOS using the Kernel Polynomial Method (KPM)~\cite{KPMReview}. The latter method allows us to iteratively evaluate LDOS via moments of the Hamiltonian and expansion into a Chebyshev polynomial basis. The accuracy of the computation can be systematically controlled by enlarging the size of the Chebyshev basis, and we have used a basis size of 400, reconstructing the final LDOS with the Jackson kernel given in Ref.~\cite{KPMReview}. We set the KPM cutoff $\epsilon{=}0.01$. We only compute the LDOS on the hole side in this case, which is confirmed to obey the expected zeroth-moment sum rule, $\int \mathrm{LDOS}_{\mathrm{KPM, hole}}(\epsilon)  \, d\epsilon  {=} \nu$~\cite{Haussmann96, MacDonald2010}. After dividing the KPM result with $\nu$, we obtain good agreement with the CF result, as seen in Fig.~\ref{fig: ED}(c).

\section{The LDOS of $\nu{=}3/7$ on the particle side}

As mentioned in the main text, we find that the LDOS on the particle side of $\nu{=}3/7$ have not converged. In this section, we discuss the origins of the lack of convergence and possible ways to obtain a converged LDOS on the particle side of $\nu{=}3/7$.

\begin{figure}[t]
    \centering    
    \includegraphics[width=0.4\linewidth]{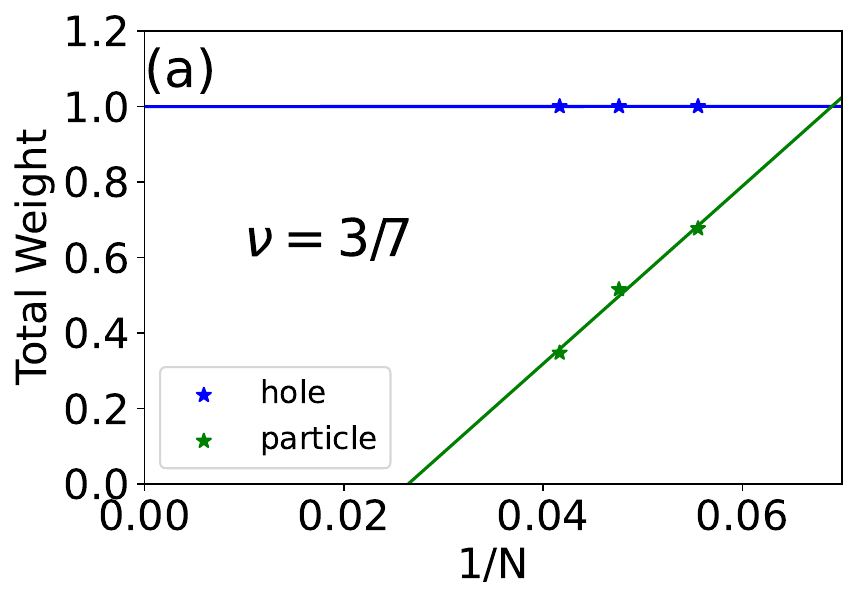}
    \includegraphics[width=0.4\linewidth]{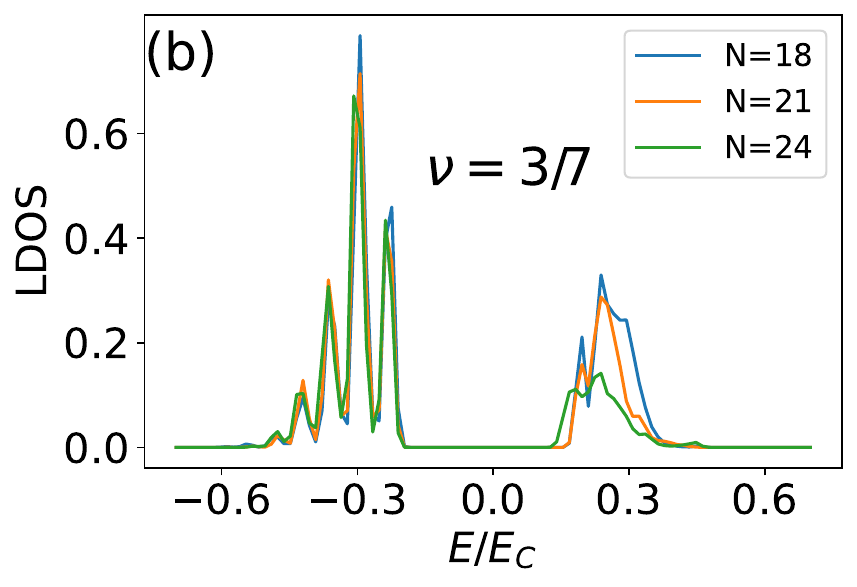}
    \caption{
     (a) The total weight $\sum_n|\eta_n^{{\pm}}|^2$ for the CF space of $\nu{=}3/7$ on both the hole and particle side. (b) LDOS for $\nu{=}3/7$. The results for the hole side were given in the main text and we reproduce them here for completeness.
     }
    \label{fig: 37par}
\end{figure}

The smallest system that has positive effective flux seen by CFs, $Q^*$, both before and after adding an electron at $\nu{=}3/7$, is $N{=}15$, $2Q{=}30$. However, after adding an electron, the state has the same electron and flux number as a ground state at $\nu{=}4/9$ and thus suffers from the ``aliasing'' problem~\cite{Ambrumenil89}. Therefore, the first system representative of 3/7 is $N{=}18$. Similar to other fillings, we keep all CF states with $E_K^*{=}0,1,2$, which include $162$ states in the $L_z{=}Q$ sector for $N{\geq} 24$ (post singular value decomposition). We show the total weight captured by the CF space and the LDOS in Fig.~\ref{fig: 37par}(a). We notice that the total weight extrapolates to a negative value in the thermodynamic limit, which is a clear signal that the $E_K^*{\leq} 2$ CF space does not cover enough portions of the Hilbert space to successfully approximate the LDOS spectrum. Fig.~\ref{fig: 37par}(b) shows the LDOS spectrum for $N{=}18,21,24$. The results of these different system sizes do not converge: even the overall shapes of the different systems visibly differ from each other.

An obvious question is what causes this convergence problem, and why it does not show up for $\nu{=}1/3, 2/5$. To answer this question, we must look at which part of the electron Hilbert space is missing in the CF space that we have retained. There are mainly two types of states. First, we ignore those CF states dressed with CF excitons. While their effective kinetic energy might be within the range $E_K^*{\leq} 2$, they have additional holes in the occupied $\Lambda$ levels of ground states. Generally, they could expand the CF space we considered, but they are unlikely to make noticeable contributions to the LDOS spectrum as their overlaps with $c_{Q}^\dagger|\Omega\rangle$ are supposed to be tiny. This is confirmed in Ref.~\cite{Gattu2023}, where these CF states dressed with CF excitons were retained but no significant difference in the LDOS spectra was observed for $\nu{=}1/3, 2/5$ on the particle side. Thereby, it is unlikely that including these states would fix the problem for the particle LDOS at $\nu{=}3/7$.

Another type of state that is absent in our CF space is that with $E_K^*{>}2$. These states become more important when the Jain sequence $\nu{=}n/(2n{+}1)$ approaches $\nu{=}1/2$, as the effective cyclotron energy $\hbar\omega_c^*$ decreases with increasing $n$. In other words, the energy difference between $E_K^*{=}2$ CF states and $E_K^*{=}3$ CF states are smaller in physical Coulomb units for $\nu{=}3/7$ than for $\nu{=}1/3,2/5$, which explains why $E_K^*{\leq} 2$ covers enough states for $\nu{=}1/3,2/5$ but not for $\nu{=}3/7$. The portion of these states in the full electron Hilbert space at a fixed filling factor increases with system size. This explains why $N{=}18,21$ show some similarity in Fig.~\ref{fig: 37par}(b) but $N{=}24$ is very different. For filling factors such as $\nu{=}4/9,5/11$, we expect this convergence problem to be more severe, as $\hbar\omega_c^*$ becomes smaller and the interval between system sizes, $2n{+}1$, becomes bigger with increasing $n$.

To solve this issue of convergence for the particle side of $\nu{=}n/(2n{+}1)$ with $n{\geq} 3$, it appears that one needs to include $E_K^*{>}2$ CF states. This is challenging to do in numerical calculation, as the dimension of CF space increases rapidly with increasing $E_K^*$. While the diagonalization of the matrix is still not a problem as the dimension is no more than a few thousand, the accurate evaluation of all the matrix elements is challenging as they all come from the Monte Carlo simulation of the CF basis states (the statistical errors stemming from Monte Carlo can aggregate). The number of the matrix elements is approximately the square of the dimension, which increases the computation task significantly even if $E_K^*$ is increased by one. Therefore, we leave the solution of this problem to future work.

\end{document}